# Differential phase shift quantum key distribution experiment over 105 km fibre

H. Takesue[1]*, E. Diamanti[2]*, T. Honjo[1], C. Langrock[2], M. M. Fejer[2], K. Inoue[1], & Y. Yamamoto[1,2]

*1. NTT Basic Research Laboratories, NTT Corporation, 3-1 Morinosato Wakamiya, Atsugi, Kanagawa, Japan.*

*2. E. L. Ginzton Laboratory, Stanford University, 450 Via Palou, Stanford, CA94305-4088, USA.*

*\* These authors contributed equally to this work.*

**Since several theoretical papers appeared in 2000[1,2], the quantum key distribution (QKD) community has been well aware that photon number splitting (PNS) attack by Eve severely limits the secure key distribution distance in BB84[3] QKD systems with Poissonian photon sources. In attempts to solve this problem, entanglement-based QKD[4-8], single-photon based QKD[9,10], and entanglement swapping-based QKD[11], have been studied in recent years. However, there are many technological difficulties that must be overcome before these schemes can become practical systems. Here we report a very simple QKD system, in which *secure* keys were generated over >100 km fibre for the first time. We used an alternative protocol of differential phase shift keying (DPSK)[12] but with a Poissonian source. We analysed the security of the DPSK protocol and showed that it is robust against hybrid attacks including collective PNS attack over consecutive pulses, intercept-and-resend (I-R) attack and beamsplitting (BS) attack, because of the non-deterministic collapse of a wavefunction in a quantum measurement.**

Although there have already been many fibre-based BB84 QKD experiments that have used Poissonian photon sources [13-20], only a few have been able to produce keys that are secure against Eve's PNS attack. The secure key generation rate of such systems scales with the square of the system transmittance, which means that long-distance secure key distribution is very difficult when using the BB84 protocol with Poissonian sources. Recently, Gobby et al. reported a secure BB84 QKD experiment over 50 km that used InGaAs APDs with a very small dark count rate[20], but with an extremely small secure key generation rate of about 0.1 bit/s. Recent studies show that the secure key distribution distance of the BB84 protocol with Poissonian sources can be significantly extended by implementing decoy states[21-23]. Although this scheme seems promising, experimental realization is still in an early stage with a reported distance of 15 km[24]. The use of a single photon source can significantly increase the secure key generation rate and distribution distance[9,10], however such a light source is not yet available for the 1.5 µm band. Entanglement-based QKD[4-8] systems are more robust against a PNS attack than BB84, but the maximum key distribution distance is limited to 30 km[8], mainly due to the difficulty involved in the generation and coincidence detection of an entangled photon pair in the 1.5 µm band. The use of quantum repeaters based on nested entanglement purification and swapping[11] constitutes another candidate for long-distance quantum communication. However, to realize such a system, we need to overcome a number of technological challenges: capturing entangled photon pairs in quantum memories either by the cavity QED technique[25] or the electromagnetically induced transparency technique[26,27], and storing qubits of information in quantum memories with a long coherence time of typically 1-10 s.

The DPSK protocol provides a simple system architecture and an effective solution to these problems. Figure 1 shows a diagram of a QKD system based on the DPSK protocol. Alice randomly modulates the phase of a weak coherent pulse train by

{0,π} for each pulse, and sends it to Bob with an average photon number of less than one per pulse. Bob measures the phase difference of each consecutive pulse with a 1-bit delay interferometer followed by two detectors placed at the interferometer output ports. Detector 1 (D1) clicks when the phase difference is 0 and detector 2 (D2) clicks when the phase difference is π. Because the average photon number per pulse is less than one, Bob observes clicks only occasionally and at a random time instance. Bob informs Alice of the time instances at which he observes clicks. From her modulation data, Alice knows which detector clicked in Bob's site. By designating D1 and D2 clicks as 0 and 1, respectively, they can share an identical bit string.

We now describe the security of the DPSK protocol. In BB84 QKD systems with Poissonian photon sources, Eve can obtain qubit information by undertaking a quantum nondemolition (QND) measurement of the photon number on each pulse and extracting one photon from a pulse containing multi-photons (PNS attack). She can also launch an I-R attack on some of the single-photon pulses and block the remaining single photon pulses. As the fibre loss increases, she can suppress more single-photon pulses. In order to overcome this hybrid attack by Eve, the average photon number per pulse must be reduced along with the fibre loss. As a result, the secure key generation rate scales with the square of the system transmittance. On the other hand, a PNS attack on each pulse is obviously useless for the DPSK protocol because information is encoded in the phase correlation between two consecutive pulses. An effective attack against the DPSK protocol is a PNS attack on two consecutive pulses, in which Eve undertakes a QND measurement of the total photon number in two consecutive pulses and extracts one photon when she observed two or more photons in two particular pulses. However, such an attack breaks the phase coherence between adjacent pulses and must induce bit errors with a probability of 1/4. Thus, a PNS attack on two consecutive pulses cannot be launched on the DPSK protocol without inducing bit errors, and so the number of bits that can be obtained by this attack is always limited by the system error rate. Eve can reduce the error probability by increasing the number of pulses for the PNS attack, but the probability that Eve obtains the same information as Bob decreases. Therefore, this collective PNS attack is not effective for DPSK protocol, even though PNS attack against BB84 protocol with a Poissonian source is very powerful.

Instead, Eve can obtain coherent copies of the quantum states of pulses by inserting a beam splitter in the transmission line. This coherent BS attack does not introduce any errors and cannot be distinguished from innocent fibre loss. Along with the BS attack, Eve can also undertake an I-R attack as long as the bit errors induced by the I-R attack are kept smaller than the innocent bit errors of the system. In the following, we discuss the security of the DPSK protocol against Eve's hybrid BS and I-R attack. Although we do not know whether this hybrid attack is the most optimal attack against DPSK protocol, this attack is more effective than the collective PNS attack on consecutive pulses that we mentioned above. A more general security analysis that is not based on specific attacks is published elsewhere.

Eve replaces a lossy fibre with transmittance $\alpha$ and imperfect detectors with a quantum efficiency $\eta$ with her lossless fibre and perfect detectors, and splits Alice's transmitter output into two paths with a BS. One beam with an average photon number of $\mu N \alpha \eta$ is sent to Bob through her lossless fibre so that Bob does not notice her eavesdropping from a change in the count rate. Here, $\mu$ is the average photon number per pulse and $N$ is the number of pulses in the coherence time of a source. The other beam with an average photon number of $\mu N(1-\alpha\eta)$ is stored in her quantum memory. After Bob announces the time instances at which he obtained clicks, Eve puts her photons into her interferometer. However, each photon in her $N$-slot wavefunction is detected completely randomly at one of $N$ different time instances. The probability that she obtains the phase modulation data at a desired time instance is $2\mu(1-\alpha\eta)$. An increase in the probability by a factor of two stems from the fact that she can use an interferometer equipped with an optical switch at the input side instead of a 50:50 BS: she turns it on only at the time instances at which she wants interferometry. Thus, for the total sifted keys of $n_{sif}$ bits, Eve has full information on $2\mu\, n_{sif}\,(1-\alpha\eta)$ bits. Note that the mutual information between Eve and Bob is independent of the system transmittance (including detector quantum efficiency) $\alpha\eta$ if $\alpha\eta \ll 1$. Therefore, the mutual information between Eve and Bob can be made small by simply choosing small $\mu$ that is independent of $\alpha\eta$ if $\alpha\eta \ll 1$. Eve can also launch an I-R attack by taking advantage of the system's innocent bit errors. Eve further splits some photons from the transmitted $\mu N \alpha\eta$ photons, and measures the phase differences with her interferometer, which is identical to Bob's. She then sends a signal only at time instances at which she detects photons. For each intercepted photon, she resends a single photon, which is split into two time slots through an interferometer identical to Bob's, in which the relative phase between the two time slots is modulated by 0 or $\pi$ according to the measurement results. When this fake photon arrives at Bob's site, he counts a photon possibly at three time instances. The ratio of the probabilities of detecting a photon at these time instances is 1:2:1, where the correct phase difference is obtained only for the second time instance. Consequently, a fake photon induces an error in the first and third time instances with a probability of 1/4. This means that Eve can attack $4en_{sif}$ photons, where $e$ represents the innocent bit error rate of the system. With these intercepted photons, Eve can obtain full information on $2en_{sif}$ bits (when Bob observes a click for the resent photons at the second time instance), and no information on the remaining bits. To summarize the above argument, the collision probability between bits owned by Bob and Eve is expressed as[1]

$$p_c = \left(\frac{1}{2}\right)^{n_{sif}\{1-2\mu(1-\alpha\eta)-2e\}} \tag{1}$$

Using the above equation, the compression factor of the privacy amplification $\tau_1$ is calculated using the following equation[1].

$$\tau_1 = 1 + \frac{1}{n_{sif}} \log_2 p_c \qquad (2)$$

Then, the secure key generation rate $R_s$ after error correction and privacy amplification is calculated to be[1]

$$\begin{aligned}R_s &= R_{ng}[1 - \tau_1 + f(e)\{e\log_2 e + (1-e)\log_2(1-e)\}] \\ &= R_{ng}[1 - 2\mu(1-\alpha\eta) - 2e + f(e)\{e\log_2 e + (1-e)\log_2(1-e)\}]\end{aligned} \qquad (3)$$

where $f(e)$ characterizes the performance of the error correction algorithm. $R_{ng}$ is the sifted key generation rate per second given by

$$R_{ng} = \mu\alpha\eta f_c \exp(-\mu\alpha\eta f_c t_d / 2). \qquad (4)$$

Here, $f_c$ and $t_d$ represent the clock frequency and the detector dead time. The factor 1/2 in the exponent arises from the fact that the average number of photons per second that reach each detector is $\mu\alpha f_c/2$. In the following theoretical calculations, we assumed a bi-directional error correction protocol. When we assume that the bit error rate $e=0$ and the transmittance $\alpha$ is small, Eq. (3) is reduced to

$$R_s \approx \mu\alpha\eta f_c (1 - 2\mu) \qquad (5)$$

Equation (5) shows that the secure key generation rate scales linearly with the system transmittance $\alpha$. This characteristic is identical to those of the coherent-state-based B92 protocol with a strong reference pulse[28] and a recently proposed protocol similar to DPSK[29].

Next we explain the up-conversion detector, which is shown schematically in Fig. 2 (a)[30]. A 1560-nm photon is combined with a strong pump light whose wavelength is 1319 nm, and injected into a PPLN waveguide. In the waveguide, a 715-nm photon is generated via the sum frequency generation (SFG) process. The internal conversion efficiency from a 1560-nm photon to a 715-nm photon exceeds 99 %. The overall conversion efficiency, including input coupling loss, waveguide loss, and output coupling loss is estimated to be approximately 65 %. The following filters suppress the noise photons such as the residual pump and SHG of the pump light. The SFG photon is detected with a single photon counting module (SPCM) based on a Si APD, which has a high quantum efficiency (about 70 %) and a small dark count rate (about 50 Hz). The overall quantum efficiency, including reflection and coupling losses and the dark count of the up-conversion detector used in our QKD experiments, is plotted in Fig. 2 (b) as a function of pump power. While the peak quantum efficiency was as high as about 37 % with a coupled pump power of around 120 mW, the dark counts increased quadratically as we increased the pump power, which was mainly due to noise photons generated by spontaneous Raman scattering process inside the waveguide and the input fibre pigtail.

The up-conversion detector for a QKD system can be operated in a non-gated mode, thanks to the low afterpulse probability of the SPCM. When we use non-gated mode detectors with a dead time $t_d$, the sifted key generation rate $R_{ng}$ of a DPSK-QKD system with equal probabilities of "0" and "π" modulation can be calculated as Eq. (4). With a small dead time and a large loss, as in our experiment, the exponential part of Eq. (4) is close to unity. Therefore, $R_{ng}$ increases with clock frequency $f_c$, which is as large as 1 GHz in our experiment. This ability of non-gated mode operation of the up-conversion detector resulted in a significant increase in the key rate in our experiment, as we describe below.

We now describe experiments that we undertook using the set-up shown in Fig. 1. At Alice's site, a continuous light from an external cavity semiconductor laser was modulated into a pulse train with a 1-GHz clock frequency using a LiNbO$_3$ intensity modulator. The pulse width was 100 ps. The phase of each pulse was then modulated by {0,π} with a LiNbO$_3$ phase modulator. After appropriate attenuation, the pulse train was sent to Bob's site through a fibre, where a 1-bit delay interferometer based on planar lightwave circuit (PLC) technology[31] was installed. The insertion loss of the interferometer was 2.5 dB, and the extinction ratio was >20 dB. The up-conversion detectors were connected to the two output ports of the interferometer. The dead time of the SPCMs was 50 ns. Each click at the photon counters was recorded using a time interval analyser (TIA). To avoid bit errors due to large dark counts introduced by the up-converter, we kept the pump power at a relatively low level. The average photon number per pulse $\mu$ was set at its optimum value for each transmittance $\alpha$, which was calculated using Eq. (3), to maximize the secure key generation rate. The optimum $\mu$ was around 0.16 - 0.18, depending on $\alpha$, $\eta$ and the dark count rate. We measured the sifted key generation rate and bit error rate five times for each transmittance, and took the average value. We calculated the secure key generation rates of the experiments with Eq. (3) using the sifted key generation rates and bit error rates obtained in the experiments. Fibre transmission experiments were undertaken using fibre spools, with Alice and Bob located in the same room. Error rate was measured by directly comparing the yielded sifted keys of Alice and Bob.

Figure 3 (a) and (b), respectively, show histograms of detected photons counted by D1 and D2 for a fixed modulation pattern after 20 km fibre transmission[32]. The tailing distribution observed in this figure was mostly due to detector timing jitter. To reduce the contribution of erroneous clicks caused by this broadening of the received signal, we applied time gating to the recorded data, which also reduces the effective dark counts per time gate. Figure 4 shows the secure key generation rate as a function of fibre length. Squares represent the secure key generation rate of fibre transmission, and x symbols show the experimental results simulating a fibre loss with an optical attenuator when we set the overall detector quantum efficiency and the time window at 8.8 % (at pump powers of approximately 15 mW) and 0.6 ns, respectively. Under these operating conditions, the total dark count rate of the two detectors was 26 kHz. As for fibre spools, we used dispersion-sifted fibre in order to avoid pulse broadening due to chromatic dispersion. The green line is the theoretical prediction for a secure key generation rate at those quantum efficiency and dark count values, where $\mu$ is optimised to maximize the secure key generation rate at each fibre loss. This theoretical curve fits very well with the experimental result. The sifted key generation rates at corresponding fibre lengths are indicated by diamonds in Fig. 4. At fibre lengths of 30 km or less, we achieved a sifted key generation rate of more than 1 Mbit/s, which is two orders of magnitude larger than the previous record[18]. The secure key generation rate was 0.455 Mbit/s over 20 km of optical fibre. Thus, even with the moderate 8.8 % quantum efficiencies, the key rate increased significantly, thanks to the non-gated mode operation of our up-conversion detectors. We then set the pump powers, the quantum efficiency, dark count, and time window width at approximately 3 mW, 2.0 %, 2.7 kHz and 0.2 ns, respectively, to further reduce the errors caused by dark counts. The experimental results are shown by a circle (105-km fibre transmission) and + symbols (attenuator), while the blue line is the theoretical calculation. By using the above detection set-up, secure key yielded with a rate of 209 bit/s. The bit error rate at 105 km was 7.95 %. The sources of errors are estimated as follows: 1 % is due to 20-dB extinction ratio of PLC interferometer, 5.5 % is from detector dark counts, and the rest is due to timing jitter. The triangles and red line show the experimental and theoretical secure key generation rates when we assumed a BB84 protocol with a Poissonian photon source and recently developed InGaAs photon counters[20]. It is clear that our result significantly outperforms a QKD system based on the BB84 protocol, both as regards secure key generation rate and distance. We also calculated the secure key distribution distance of a BB84 QKD system with a Poissonian source combined with our up-conversion detectors operated in the same condition as in our 105 km transmission experiment (i.e. 2.0-% quantum efficiency, 1.35-kHz dark count rate per detector, and 0.2-ns time window), using theory described in Ref. 2. As a result, the secure key distribution distance is at most 58 km over fibre with 0.2 dB/km loss even when we can eliminate all the optical loss except fibre. This clearly shows that the use of the DPSK protocol was essential for enlarging secure key distribution distance up to 105 km.

In our experiment, the secure key distribution distance was limited by two impairments of the up-conversion detectors: a large dark count rate resulting from noise photons generated in the waveguide, and the large timing jitter of the SPCM[33]. If we can eliminate the noise photons[34] and make the timing jitter negligibly small compared with pulse width, the secure key distribution distance will reach 300 km. Therefore, the development of entanglement-based quantum repeater systems would be meaningful only when the system length can exceed 300 km.

relatively low loss point. As a result, the maximum sifted key generation rate did not greatly exceed 1 Mbit/s. The large timing jitter also led to a larger bit error rate, which resulted in a reduction of the secure key distribution distance.

34. For example, noise photons due to spontaneous Raman scattering are expected to be suppressed by setting pump wavelength longer than signal wavelength.

The authors would like to thank N. Lutkenhaus, M. Koashi and E. Waks for discussion on security issues. They also wish to thank M. Morita and Y. Tokura for encouragements throughout this research.

Correspondence and requests for materials should be addressed to H. T. (e-mail: htakesue@will.brl.ntt.co.jp), or E. D. (e-mail: ediam@stanford.edu)

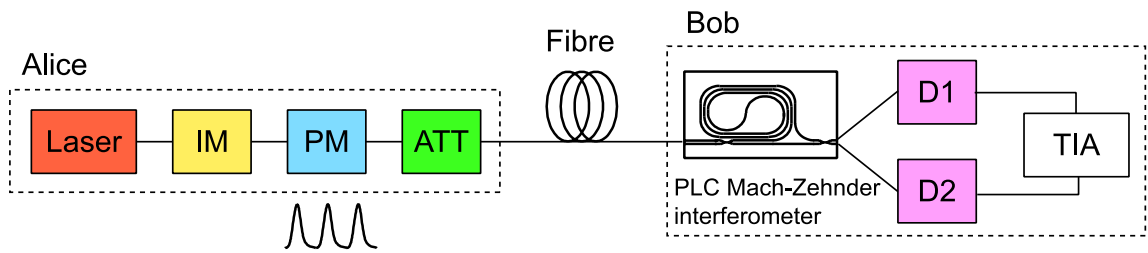

Figure 1: Schematic diagram of DPSK-QKD. IM: intensity modulator, PM: phase modulator, ATT: optical attenuator, PLC: planar lightwave circuit, D1, D2: detectors, TIA: time interval analyser.

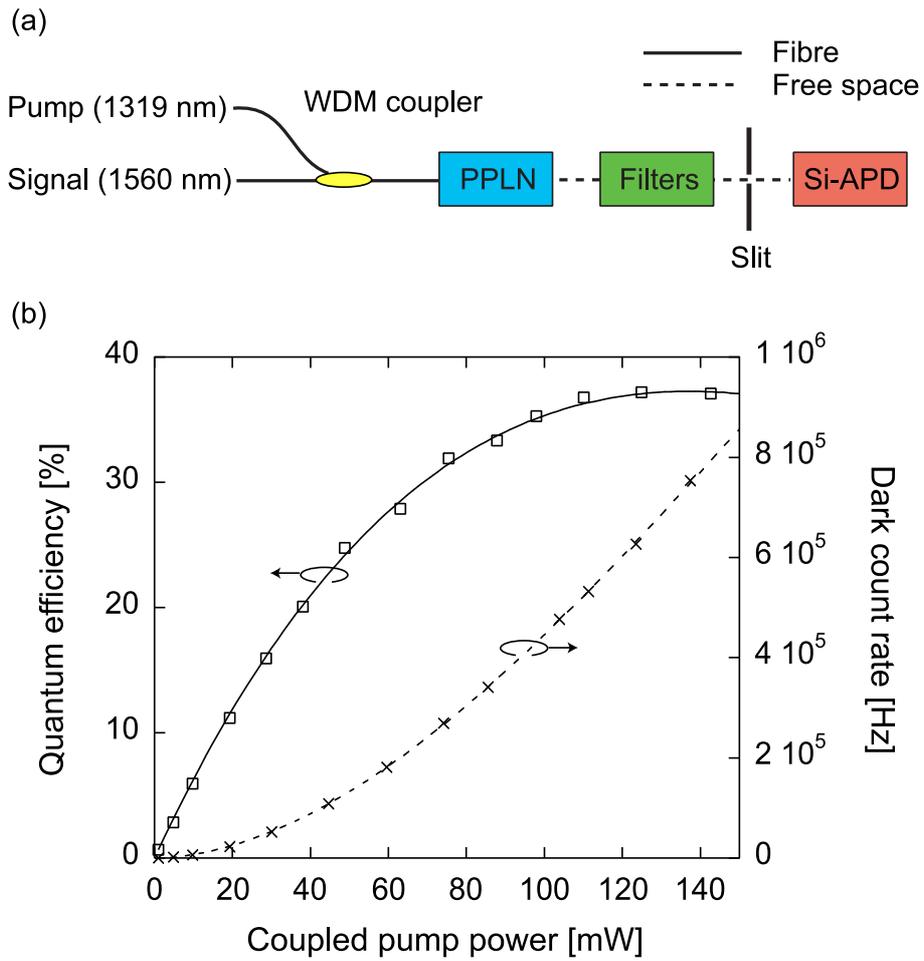

Figure 2: (a) Schematic diagram of up-conversion detector. (b) Quantum efficiency and dark count rate as functions of pump power.

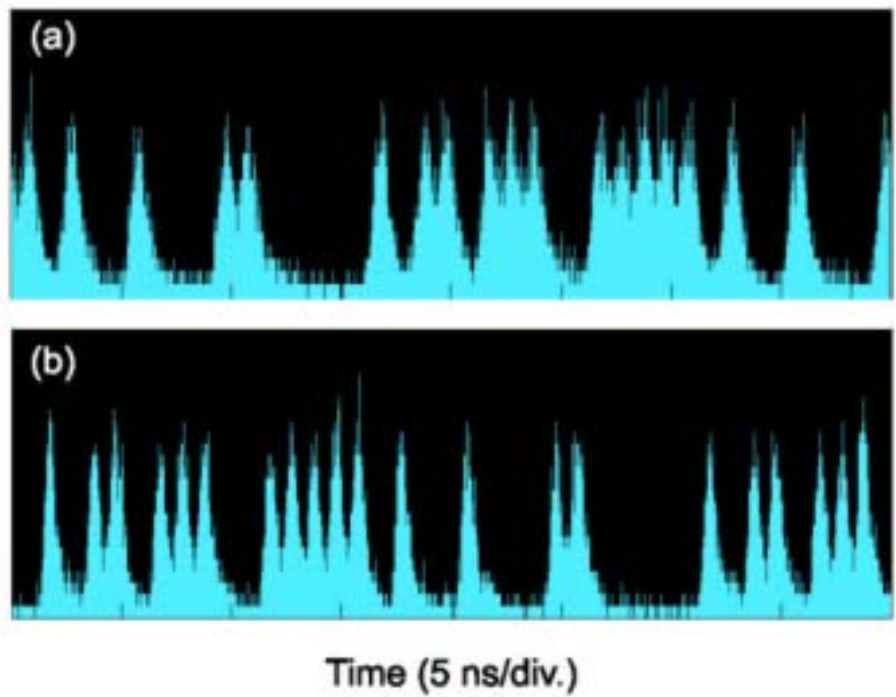

Figure 3: Histograms of the received signal at (a) D1 and (b) D2 for a fixed modulation pattern.

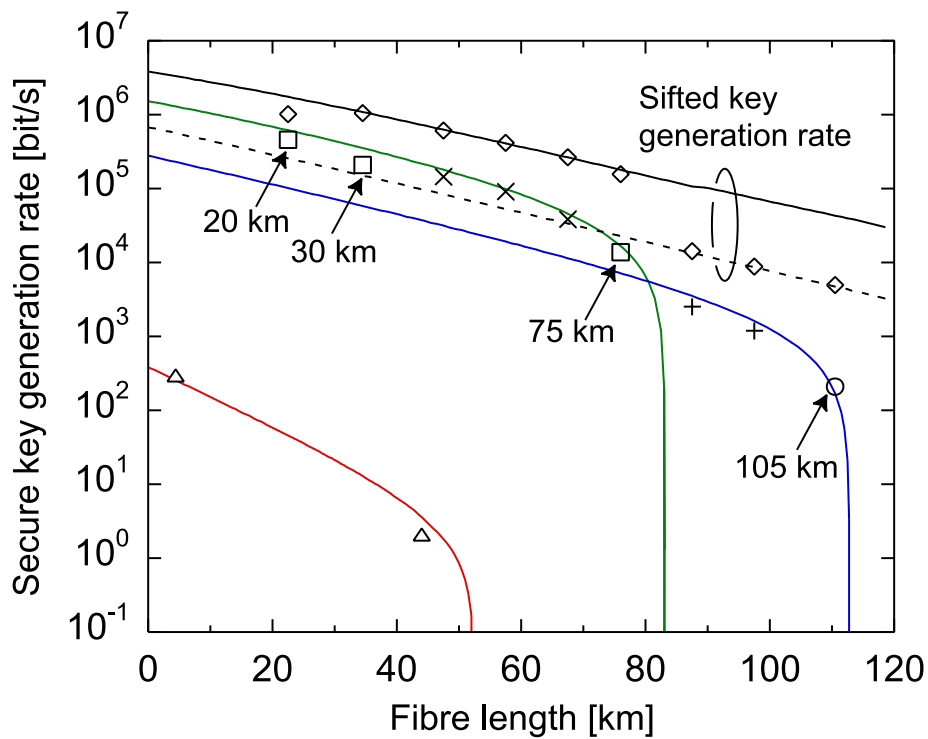

Figure 4: Secure key generation rate as a function of length of fibre with 0.2 dB/km loss. Squares: fibre transmission (experiment, $\eta$ =8.8 %), x: simulated points with attenuator (experiment, $\eta$ =8.8 %). Circle: fibre transmission (experiment, $\eta$ =2.0 %). +: simulated points with attenuator (experiment, $\eta$ =2.0 %). Triangles: fibre transmission with BB84 and InGaAs APD (experiment described in [19]). Green line: DPSK and up-conversion detector operated at $\eta$ =8.8 % (theory). Blue line: DPSK and up-conversion detector operated at $\eta$ =2.0 % (theory). Red line: BB84 and InGaAs APD (theory). 3 % system error is assumed. The characteristic of InGaAs APD is based on [19]. Diamonds show the experimental data for sifted key generation rates at corresponding fiber lengths. Solid and dotted black lines shows the theoretical prediction of sifted key generation rates for $\eta$=8.8 and 2 %, respectively.